\documentclass[aps,prl,twocolumn,groupedaddress]{revtex4-1}
\usepackage{graphicx}
\usepackage{amsfonts}
\usepackage{amsmath}
\usepackage{color}
\usepackage{soul}
\usepackage{booktabs}

\def\opone{\leavevmode\hbox{\small1\kern-3.8pt\normalsize1}}

\begin{document}

\title{Demonstration of analyzers for multimode photonic time-bin qubits}

\author{Jeongwan Jin$^{1,2}$} \altaffiliation{Present address: National Research Council of Canada, 1200 Montreal Road, Ottawa, Ontario K1A 0R6, Canada} \email{jeongwan.jin@nrc-cnrc.gc.ca}
\author{Sascha Agne$^{1,2}$} \altaffiliation{Present address: Rockefeller University, 1230 York Avenue, New York, New York 10065, USA}
\author{Jean-Philippe Bourgoin$^{1,2}$}
\author{Yanbao Zhang$^{1,2}$} \altaffiliation{Present address: NTT Basic Research Laboratories, Morinosato-Wakamiya, Atsugi, Kanagawa 243-0198, Japan}
\author{Norbert L\"{u}tkenhaus$^{1,2}$}
\author{Thomas Jennewein$^{1,2,3}$} \email{thomas.jennewein@uwaterloo.ca}

\affiliation{$^1$ Institute for Quantum Computing, University of Waterloo, 200 University Avenue West, Waterloo, Ontario N2L 3G1, Canada \\ $^2$ Department of Physics and Astronomy, University of Waterloo, 200 University Avenue West, Waterloo, Ontario N2L 3G1, Canada \\$^3$ Quantum Information Science Program, Canadian Institute for Advanced Research, Toronto, Ontario M5G 1Z8, Canada}

\date{\today}

\begin{abstract}
We demonstrate two approaches for unbalanced interferometers as time-bin qubit analyzers for quantum communication, robust against mode distortions and polarization effects as expected from free-space quantum communication systems including wavefront deformations, path fluctuations, pointing errors, and optical elements. Despite strong spatial and temporal distortions of the optical mode of a time-bin qubit, entangled with a separate polarization qubit, we verify entanglement using the Negative Partial Transpose, with the measured visibility of up to $0.85 \pm0.01$. The robustness of the analyzers is further demonstrated for various angles of incidence up to 0.2$^{\circ}$. The output of the interferometers is coupled into multimode fiber yielding a high system throughput of 0.74. Therefore, these analyzers are suitable and efficient for quantum communication over multimode optical channels.

\end{abstract}


\maketitle

\section{Introduction}

Quantum communication experiments in free space~\cite{R. Ursin 2007, J. Yin 2017, H. Takenaka 2017, S. Nauerth 2013, C. Pugh 2017} are usually based on polarization-encoded photons due to their robustness against atmospheric turbulence~\cite{D. Hohn 1969}. However, the quality of polarization is fundamentally limited by nonideal steering optics and telescopes, and reference-frame alignment~\cite{M. Jofre 2012, C. Bonato 2006}. For instance, photon polarization typically experiences phase shifts when reflected off optical surfaces with a nonzero angle of incidence (AOI), leading to errors in encoded information. This problem becomes significant when communicating parties are located on moving platforms, such as aircrafts or satellites, where a signal-tracking system introduces fluctuations in the system alignment~\cite{C. Bonato 2006}. Moreover, polarization can be changed in a nonunitary manner that cannot be corrected when passing through tempered glasses and polycarbonates, which are widely used in vehicles and buildings. 

Time-bin encoding~\cite{J. Brendel 1999} is an interesting alternative due to its immunity against polarization drifts. The method has been demonstrated with various quantum communication protocols in optical fibers, including plug-and-play~\cite{A. Muller 1997, D. Stucki 2002}, differential phase shift~\cite{K. Inoue 2002}, and coherent one-way~\cite{D. Stucki 2009} quantum key distribution (QKD) protocols, as well as Mach-Zehnder interferometer-based systems~\cite{P. D. Townsend 1993}, quantum teleportation~\cite{I. Marcikic 2003}, and elements of quantum repeaters~\cite{E. Saglamyurek 2011}. Despite its versatility, time-bin encoding has been implemented only in single-mode optical fibers and is generally considered impractical for free-space channels. The reason is that spatial and temporal modes of a photon are distorted during the transmission through multimode optical channels such as multimode fibers and turbulent free space~\cite{D. L. Fried 1992}. Mode distortions introduce path distinguishabilities in unbalanced Michelson or Mach-Zehnder interferometers, which are typically used as time-bin analyzers, hindering single-photon interference required for analysis of time-bin states. In addition, telescope pointing errors as well as turbulence-induced angular fluctuations~\cite{L. Karl 2006} degrade the quality of interference even further. For example, Ursin {\it et al.}~\cite{R. Ursin 2007} reported turbulence-induced AOI errors of up to 4 $\times10^{-3}$ degrees over a horizontal 143 km link, and we reported~\cite{J. -P. Bourgoin 2015} pointing errors of 6 $\times 10^{-2}$ degrees on a moving platform. This leads to phase shift and visibility reduction. Spatial filters such as single-mode fibers can be used to combat those problems, which, however, discards most of the impinging photons~\cite{H. Takenaka 2012}.

Here we investigate two types of unbalanced Michelson interferometers for analyzing time-bin qubits encoded on spatially and temporally distorted photons, originally developed for Doppler spectroscopy of stars ~\cite{G. A. Vanasse 1977}. These so-called field-widened interferometers use imaging optics, or carefully chosen refractive indices, to correct AOI-induced phase shifts and visibility reduction, hence achieving a larger field of view than conventional Michelson interferometers. However, it was not known whether such interferometers are capable of analyzing entanglement or quantum sates transmitted over multimode optical channels. To prove that, we take the following two steps. First, we compare the performance between the conventional and field-widened interferometers using classical light. Next, by utilizing quantum entanglement, we demonstrate the viability of the interferometers as multimode time-bin receivers for quantum applications. 

\section{Multimode Time-Bin Analyzer Methods}

Let us consider an unbalanced Michelson interferometer with long and short paths of lengths $l_{\rm{L}}$ and $l_{\rm{S}}$, respectively. While the path-length difference for zero-angle incidence is simply $\Delta l_0=2(l_{\rm{L}}-l_{\rm{S}})$, a nonzero AOI translates into an angle-dependent path length and a lateral offset as the beam propagates. Using geometrical ray tracing through the interferometer, we find that the path-length difference is given by

\begin{multline}
\Delta l(\alpha) = \frac{\Delta l_0}{2}\left(\frac{1}{\cos(\alpha)} +\frac{1-\tan(\alpha)}{\cos(\alpha)+\sin(\alpha)}\right)\\
+\delta(\alpha)\tan\left(\alpha-\frac{\pi}{4}\right),
\label{eq:pathlengthdifference}
\end{multline}

\noindent where $\delta(\alpha)=\Delta l_0\tan(\alpha)/(1+\tan(\alpha))$ is the lateral offset between the two rays coming from each path of the interferometer at the output beam splitter [see Fig. 1]. From Eq. (1), we see that a nonzero AOI introduces path distinguishability and rapidly modulates the interferometer phase at the same time. The relative phase between the two paths is very sensitive to the AOI, with a predicted $\pi$--shift per $2 \times10^{-5}$ degrees input angle variation. In order to quantify interference degradation due to input-angle fluctuations, we compute the interference visibility. Considering a single-mode Gaussian beam with intensity $I_0$ and a beam width $\sigma$ at the interferometer input, the visibility is given by~\cite{H. Mishina 1974}

\begin{figure}
\includegraphics[width=\linewidth]{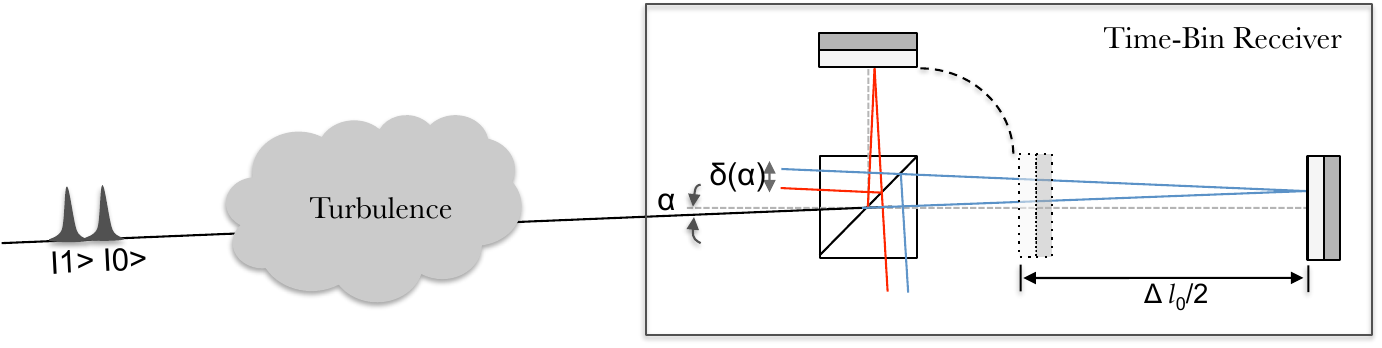}
\caption[entanglementverification]{(Color online) \textbf{Time-bin-based quantum communication in a turbulent free-space channel.} When a time-bin-encoded photon, whose path is deviated by atmospheric turbulence as well as telescopes misalignment, enters a time-bin receiver with variable angle of incidence $\alpha$, a lateral offset $\delta(\alpha)$ occurs between the paths (red and blue line) at the interferometer exit. This is due to the receiver length asymmetry and reduces the interference quality, resulting in lower distinguishability of the time-bin states in superposition bases. Turbulence-induced spatial-mode distortions further lower the interference visibility [see text for details].}
\label{concept}
\end{figure} 

\begin{equation}
\mathcal{V}(\alpha) = \mathcal{V}_0\;{\rm exp}\left( - \left(\frac{\Delta l_0 \tan (\alpha)}{ \sqrt{2}\sigma(1+\tan(\alpha))}\right)^2\right), 
\label{eq:visibility}
\end{equation} 

\noindent where $\mathcal{V}_0$ denotes the system visibility at zero AOI. For instance, with $\sigma = 1.49$\,mm and  $\Delta l_0 = 0.60$~m, due to Eq. (2), the visibility will drop to 0.70 for $\alpha$ = 0.1$^{\circ}$ and $\mathcal{V}_0$ = 0.91. The relationship Eq. (2) is verified experimentally with a single-mode beam [see Fig. 2(a)], generated by a continuous-wave laser at 776\,nm. For instance, as shown in Fig. 2(d), the initial interference visibility of $\mathcal{V}_0^{\rm{single}}$ = 0.91~$\pm$~0.01 decreases rapidly with AOI. The same laser beam is then sent through a multimode fiber, thereby distorting it into a multimodal beam~\cite{I. N. Papadopoulos 2012}  which mimicks the effect of turbulent atmosphere [Fig. 2(b); see~\cite{R. Ursin 2007, D. L. Fried 1992} for comparison]. Despite lengthy and careful alignment we were able to obtain only a maximum visibility of $\mathcal{V}_0^{\rm{multi}}$ = 0.16~$\pm$~0.01, which, as shown in Fig. 2(e), drops to zero with an AOI of 0.2$^{\circ}$. These observations clearly show that, given the expected angular deviations reported for free-space quantum channels, it would be technically very challenging to achieve a reliable, stable and efficient operation of time-bin qubit analyzers using standard interferometers. \\

\begin{figure*}
\includegraphics[height=10cm, width=15cm]{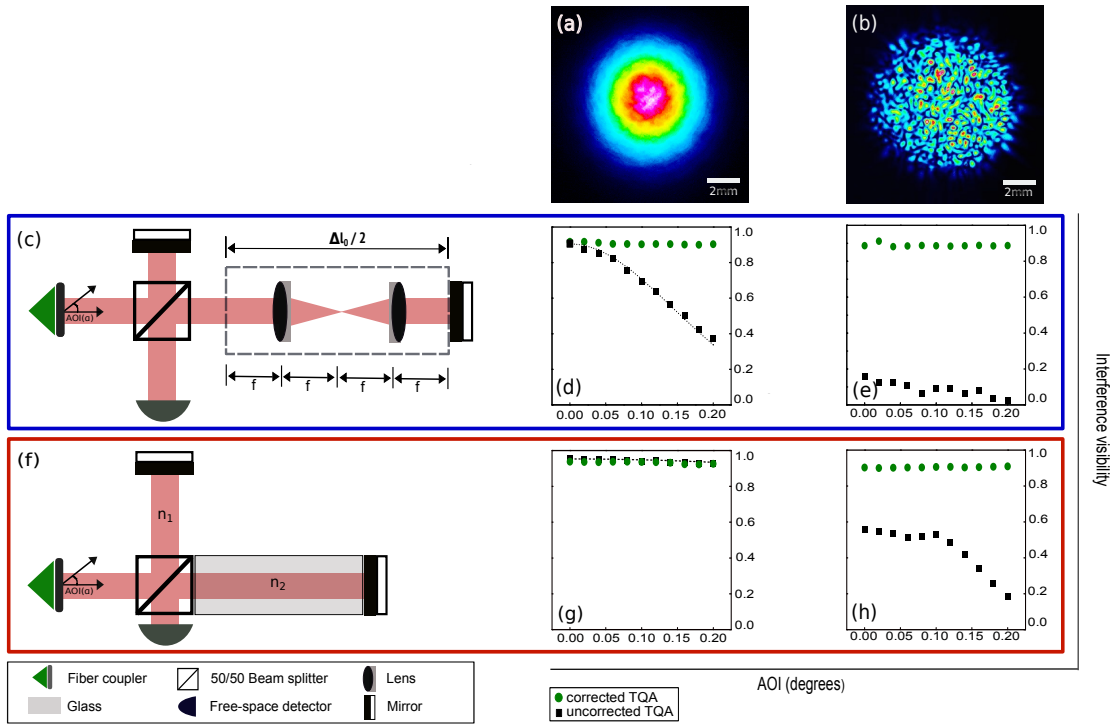}
\begin{center}
\caption[timebinqubitanalyzer]{(Color online) \textbf{Our multimode time-bin qubit analyzers (MM-TQA).} \textbf{(a)} Image of the incident single-mode Gaussian beam, captured with a beam-profiling camera (WinCamD-UCD12). \textbf{(b)} Image of the incident multimode beam, generated by a 1\,m-long step-index multimode fiber (Thorlabs M43L01). \textbf{(c)} Schematic diagram of our MM-TQA using imaging optics. Temporal separation between paths is set to 2.0\,ns ($\Delta l_0 = 0.60~\rm{m}$). $\rm{f}$ denotes a focal length of the lens. \textbf{(d)} Interference visibility for the single-mode beam with (green circles) and without (black squares) relay optics. Measured visibilities are in good agreement with theoretical prediction of Eq. (2) (dashed black line). \textbf{(e)} Interference visibility for the multimode beam with (green circles) and without (black squares) relay optics. \textbf{(f)} Schematic diagram of our MM-TQA with different refractive indices for each path. Temporal separation between paths is set to 0.57\,ns ($\Delta l_0 = 0.17~\rm{m}$). \textbf{(g)} Interference visibility for the single-mode beam with (green circles) and without (black squares) glass. Measured visibilities are in good agreement with theoretical prediction of Eq. (2) (dashed black line). \textbf{(h)} Interference visibility for the multimode beam with (green circles) and without (black squares) glass. The uncorrected TQA visibilities in (g) and (h) are higher than in (d) and (e), because shorter path-length difference introduces less path distinguishability. Uncertainties are smaller than symbol size. }
\end{center}
\label{analyzers}
\end{figure*} 
 
These interference challenges are overcome by utilizing relay optics in the long arm of the unbalanced Michelson interferometer (Method 1). The idea is to reverse differences in the evolution of spatial modes over the length $\Delta l$ in the long arm, as shown in Fig. 2(c). This effectively guarantees identical wavefront evolutions in the short and long paths of the interferometer. Consequently, spatial indistinguishability is restored regardless of spatial mode and AOI of the input beam. For verification, we set $\Delta l$ = 0.60\,m (2.0\,ns) and measure interference visibilities by applying voltages to a piezo mounted on a mirror in the short path, allowing it to change the phase of the interferometer at various AOIs. Having a single-mode beam as an input, we obtain an interference visibility of $\mathcal{V}^{\rm{single}}$= 0.91~$\pm$~0.01, which remains constant as the AOI is varied [see Fig. 2(d)]. The visibility and error are extracted from a sinusoidal fit of measured data. The improvement is further confirmed by measurements with a multimode beam [Fig. 2(b)] where the high visibility of $\mathcal{V}^{\rm{multi}}$=0.89~$\pm$~0.01 [Fig. 2(e)] demonstrates that the interferometer design is robust against highly distorted beams. This is noteworthy as it allows us to couple the output of the interferometer into a multimode fiber, yielding a high coupling efficiency of 0.87 for delivery of photons to the detector. Phase-recovery capacity is discussed in the section Measurements and Results.\\

\begin{figure*}
\includegraphics[height=10cm, width=15cm]{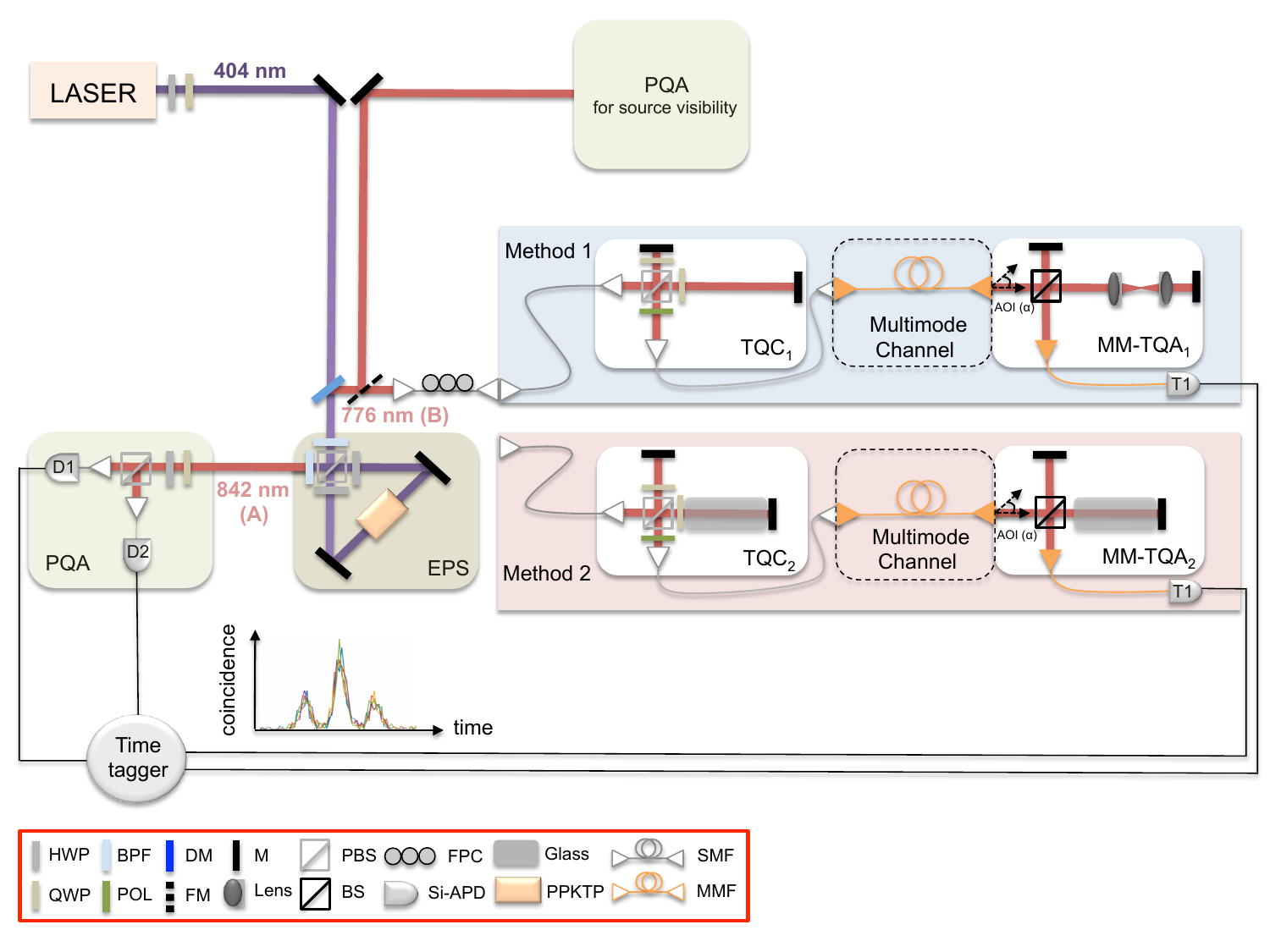}
\caption[experimentalsetup]{(Color online) \textbf{Experimental setup.} The polarization-entangled photon-pair source (EPS) is described in the text.  For projection measurements, photon A is directed to a polarization-qubit analyzer (PQA), consisting of a quarter-wave plate (QWP), a half-wave plate (HWP), a polarizing beam splitter (PBS), and silicon avalanche photodiodes (Si-APDs).  After reflection at a dichroic mirror (DM), via a flip mirror (FM), photon B is sent either to a PQA or a time-bin qubit converter (TQC$_{1(2)}$) followed by a multimode fiber  (MMF) and a multimode time-bin qubit analyzer (MM-TQA$_{1(2)}$). All detection signals are sent to a time tagger for data analysis.}
\label{experiment}
\end{figure*} 

The second type of interferometer we study is based on the use of media with different refractive indices for the paths of the unbalanced interferometer (Method 2), as shown in Fig. 2(f). The combination of glass and mirror produces a virtual mirror situated closer 
to the interferometer beam splitter. With the appropriate choice of refractive index and glass length, we can match the distance beam-splitter-to-virtual-mirror to the corresponding distance of the real mirror in the short arm. This effectively balances the interferometer. More specifically, let us consider the situation in which an input beam enters the interferometer with an angle of $\alpha$. The optical path difference in the interferometer is given by $\Delta l = 2 (n_{\rm{L}} l_{\rm{L}} cos \alpha_{\rm{L}} - n_{\rm{S}} l_{\rm{S}} cos \alpha_{\rm{S}})$, where $n_{\rm{L(S)}}$ and  $\alpha_{\rm{L(S)}}$ denote refractive index and reflection angle from a mirror in path $l_{\rm{L(S)}}$, respectively. Using Snell's law and Taylor's expansion, the difference is approximated as $2 (n_{\rm{L}} l_{\rm{L}} - n_{\rm{S}} l_{\rm{S}}) - \sin^2\alpha(l_{\rm{L}}/n_{\rm{L}}-l_{\rm{S}}/n_{\rm{S}})$ for small angles $\alpha_{\rm{L}}$ and $\alpha_{\rm{S}}$. With a proper choice of refractive indices for both paths, we can remove the second term so that $\Delta l$ becomes insensitive to AOI, thus restoring indistinguishability at the interferometer output. In our implementation, we use 118\,mm-long glass with the refractive index $n$=1.4825 in the long path and none in the short path, providing an optical path-length difference of $\Delta l $  = 0.17\,m (0.57\,ns). Interference visibilities of 0.94~$\pm$~0.01 [see Fig. 2(g)] and 0.90~$\pm$~0.01 [see Fig. 2(h)] are measured with a single-mode and multimode beam, respectively, which remain constant as the AOI is varied. Hence, correcting optics not only improves performance at higher AOI but is also necessary to enable high interference visibility with a multimode beam. 

\section{Quantum Communication Experiments}

We demonstrate the viability of our MM-TQAs for use with quantum signals using the experimental setup depicted in Fig. 3. Light from a 404\,nm continuous-wave laser with an average power of 6\,mW pumps a periodically poled potassium titanyl phosphate crystal inside a Sagnac interferometer. This generates polarization entangled photon pairs at 842\,nm (A) and 776\,nm (B) in a form of $|\psi\rangle = \frac{1}{\sqrt{2}} (|\rm{H\rangle_A |V}\rangle_B +|\rm{V\rangle_A |H}\rangle_B)$ via type-II spontaneous parametric down-conversion. Here, $|\rm{H\rangle}$ and $|\rm{V\rangle}$ are the horizontal and vertical polarization states, respectively, forming the eigenstates of the computational basis. Unused pump photons are removed by band-pass filters. While photon A is directed to a polarization analyzer (PQA), photon B is sent either to a separate PQA or a time-bin converter (TQC) followed by a multimode channel and a MM-TQA for various measurements. The PQAs measure reference entanglement visibility with the source of polarization entanglement.

\begin{figure*}
\includegraphics[width=\textwidth]{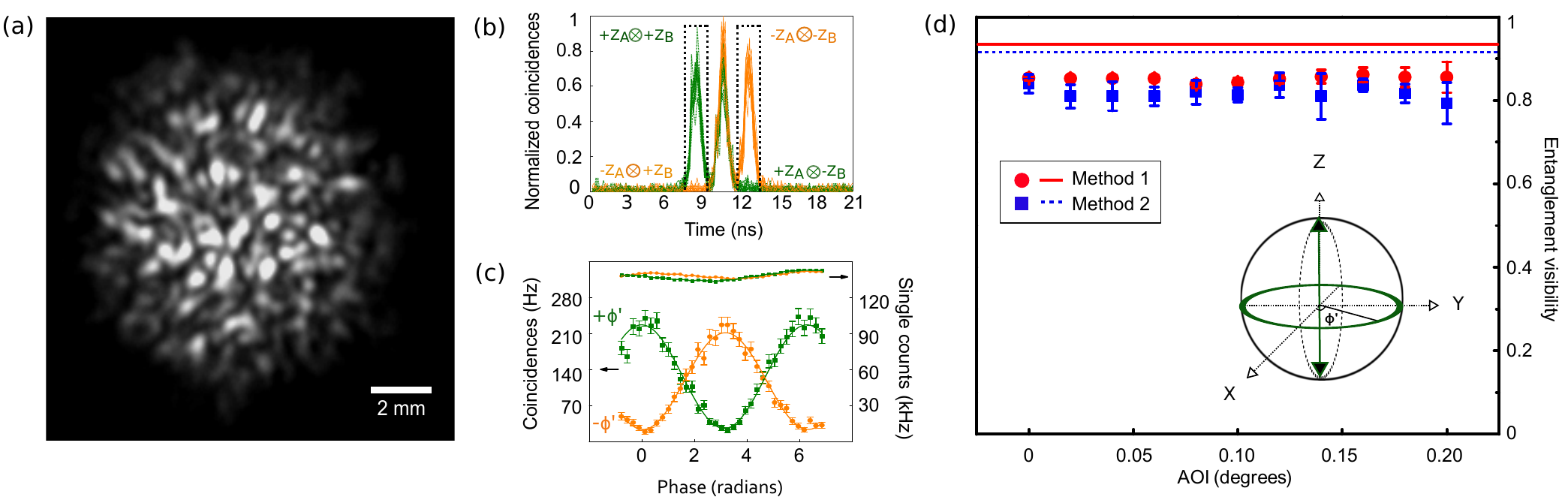}
\caption[entanglementvisibility]{(Color online) \textbf{Experimental results for entangled photons analyzed with our MM-TQAs.} \textbf{(a)} Spatial mode of photon A before entering the MM-TQA. The image is captured with an electron multiplier CCD camera (Hamamatsu C9100-13). \textbf{(b)} Joint detections for the projection $+ \rm{Z_A} \otimes \pm \rm{Z_B}$ (dotted green lines) and $- \rm{Z_A} \otimes \pm \rm{Z_B}$ (solid orange lines) as a function of detection-time difference between the photon A and B. \textbf{(c)} Joint detections for the projection $+ \phi_{\rm{A}^{\prime}} \otimes \phi_{\rm{B}}$ (green squares) and $- \phi_{\rm{A}^{\prime}} \otimes \phi_{\rm{B}}$ (orange circles) as a function of phase $\phi_{\rm{A}^{\prime}}$ of a polarization qubit using motorized wave plates. Visibilities $\mathcal{V}_{\pm \phi}$ are obtained from sinusoidal fittings. Single counts remain essentially constant as we scan the phase. \textbf{(d)} The measurements (b) and (c) are repeated for different AOIs. Red circles and blue squares are average visibilities obtained with Method 1 and 2, respectively. Solid red and dotted blue lines are reference visibilities measured directly with our source of polarization-polarization entanglement prior to the measurement of time-polarization entanglement using Method 1 and 2, respectively. We maintain high entanglement visibility (close to source visibility) despite the high multimode nature of incoming photons.} 
\label{result}
\end{figure*} 

To convert the polarization state of photon B into a time-bin state, we use an unbalanced interferometer as a TQC$_{1(2)}$~\cite{F. Bussieres 2010}, whose path-length difference is matched to the MM-TQA$_{1(2)}$. At the input polarizing beam splitter of the TQCs, a photon is either reflected or transmitted into the short or long path, respectively. A fiber-polarization controller (FPC) ensures the faithful mapping of the vertical (horizontal) polarization onto the early (late) temporal bin. The inserted quarter-wave plate in each path guides photons to the desired output port. Leaving the TQCs, photons pass through a polarizer set to an equal superposition between the polarizations, erasing polarization information for each time-bin state at the cost of 50 \% transmission loss. This completes the map $|\rm{V}\rangle \mapsto |\rm{E}\rangle$ and $|\rm{H}\rangle \mapsto |\rm{L}\rangle$, resulting the two-photon entangled state in a form of 

\begin{equation}
|\psi\rangle = \frac{1}{\sqrt{2}}(|\rm{H}\rangle_A|\rm{E}\rangle_B + e^{i \phi}|\rm{V}\rangle_A|\rm{L}\rangle_B),
\label{eq:state}
\end{equation} 

\noindent where $|$E$\rangle$ ($|$L$\rangle$) denotes the quantum state in which photon B is in early (late) temporal mode, and $\phi$ is a relative phase between the modes introduced during the conversion process. Photon B then travels through a 1m-long step-index multimode fiber, as a multimode channel, distorting the spatial mode [see Fig. 4(a)] and temporal mode (measured dispersion is about 50 ps, drastically exceeding the photon's coherence time of 3.2\,ps~\cite{D. R. Hamel 2010}), prior to entering the MM-TQAs. For Method 2, we calculate for the glass a dispersion of 5.48 waves/nm and 5.21 waves/ $^{\circ}$C. In order to minimize dispersion effects, we symmetrize the paths in the time-bin converter and analyzer. After being analyzed in the MM-TQAs, both photons A and B are detected by silicon avalanche photodiodes and the detection signals are sent to a time tagger and computer for data analysis. 

\section{Measurements and Results}

The MM-TQA performance is verified by entanglement visibility measurements. For the measurements, photons A and B are directed to a polarization and a time-bin qubit analyzer, respectively. Each qubit is first projected onto the computational basis, i.e., $|\pm \rm{Z}\rangle\langle\pm \rm{Z}|$, where $|+ \rm{Z_A}\rangle \equiv |\rm{H}\rangle$, $|- \rm{Z_A}\rangle \equiv |\rm{V}\rangle$, $|+ \rm{Z_B}\rangle \equiv |\rm{E}\rangle$, and $|- \rm{Z_B}\rangle \equiv |\rm{L}\rangle$. The coincidence counts are used to calculate correlation visibilities $\mathcal{V}_{\pm \rm{Z},\pm \rm{Z}} \equiv (\rm{N}_{\pm \rm{Z} \pm \rm{Z}}-\rm{N}_{\mp \rm{Z} \pm \rm{Z}}) / (\rm{N}_{\pm Z \pm Z}+\rm{N}_{\mp Z \pm Z})$, from which we obtain the average $\mathcal{V}_{\rm{Z}}=(\mathcal{V}_{\rm{+Z},\rm{+Z}} + \mathcal{V}_{\rm{-Z},\rm{-Z}})/2$. Here, $\rm{N}_{\rm{ij}}$ denotes the joint-detection counts when polarization qubit A is projected onto $|\rm{i}\rangle\langle \rm{i}|$ and time-bin qubit B onto $|\rm{j}\rangle\langle \rm{j}|$, where i, j $\in$ \{$+\rm{Z}, -\rm{Z}$\} [see Fig. 4(b)]. The qubits are then projected onto superposition states, i.e., $|\pm \phi\rangle \langle \pm \phi |$, where $|\pm \phi_{\rm{A(B)}}\rangle \equiv \frac{1}{\sqrt{2}} (|+\rm{Z_{\rm{A(B)}}}\rangle \pm e^{i \phi_{\rm{A(B)}}} |-\rm{Z_{\rm{A(B)}}}\rangle)$. To measure the visibility, we vary the relative phase between basis states of the polarization qubit. A complete scan of the phase along the $\rm{XY}$-plane of the Bloch sphere is performed [see Fig. 4(c)], yielding the average $\mathcal{V}_{\rm{\phi}} = (\mathcal{V}_{\rm{+\phi}} + \mathcal{V}_{\rm{-\phi}})/2$. These allow us to compute an average visibility $\mathcal{V}_{\rm{avg}}\equiv\mathcal{V}_Z/3 + 2\mathcal{V}_{\phi}/3$. For a concluding assessment of the performance of the MM-TQAs, we compare $\mathcal{V}_{\rm{avg}}$ to the source visibility obtained from the original polarization entanglement. This is done by routing photon B to a polarization analyzer. For Method 1(2), we measure visibilities of $\mathcal{V}_Z$=0.95~$\pm$~0.01 (0.92~$\pm$~0.01) and $\mathcal{V}_{\phi}$=0.80~$\pm$~0.01 (0.77~$\pm$~0.01), yielding an average visibility of $\mathcal{V}_{\rm{avg}}$= 0.85~$\pm$~0.01 (0.82~$\pm$~0.01). The difference to the source visibility of 0.93~$\pm$~0.01 (0.91~$\pm$~0.01), as shown in Fig. 4(d), mainly stems from the nonunity interference visibilities of our MM-TQAs.

\begin{figure}
\includegraphics[width=\linewidth]{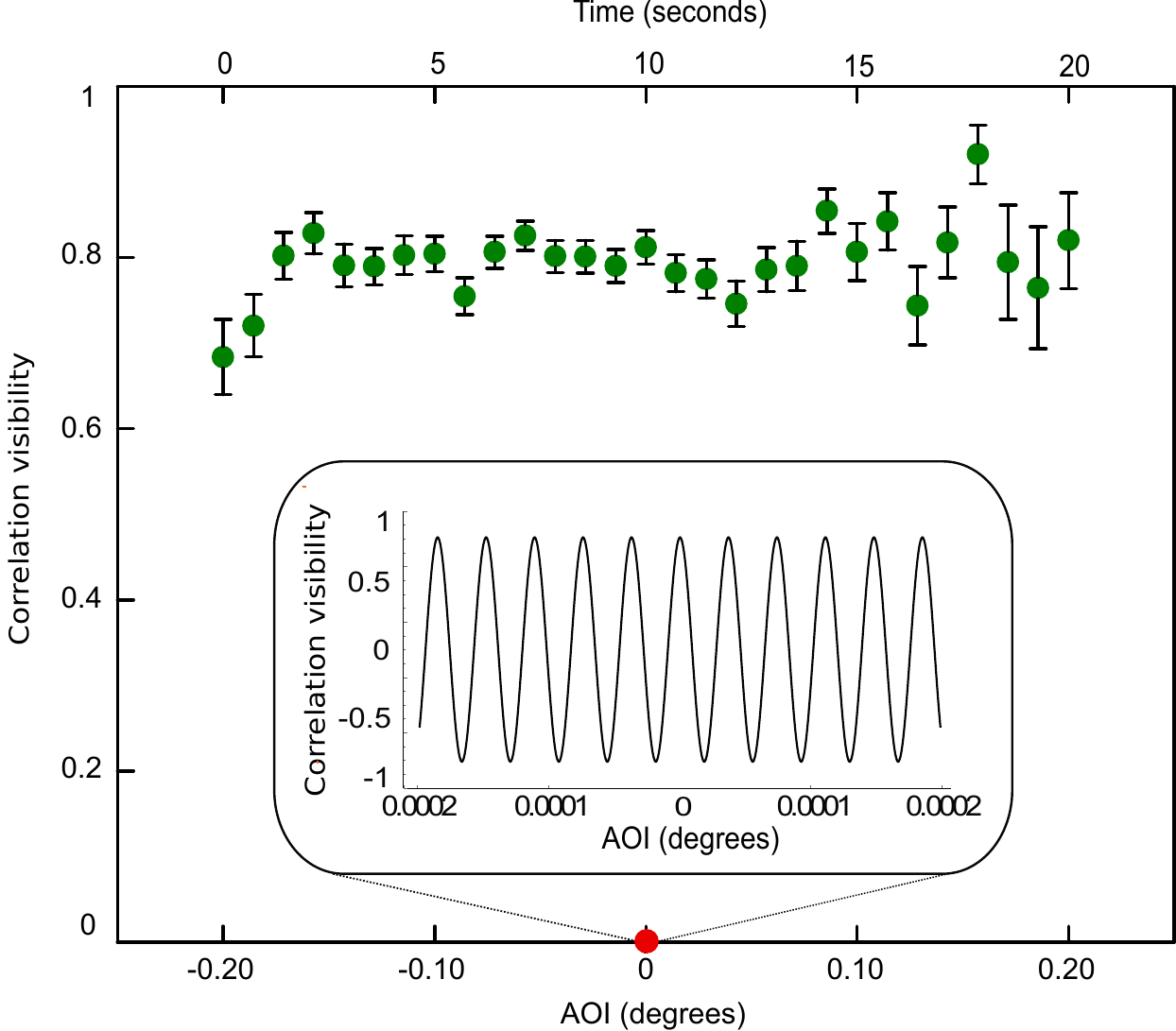}
\caption[expectationangle]{(Color online) \textbf{Phase stability of our MM-TQA} Correlation visibilities $\mathcal{V_{\rm{\phi_{A^{\prime}}}, \rm{\phi_B}}}$ (green circles) are measured using Method 1 as the AOI is continuously varied from -0.2$^{\circ}$ to +0.2$^{\circ}$ over 20 sec. The inset shows the calculated visibilities without relay optics as a function of AOI. Due to AOI-induced phase fluctuations,  the value rapidly changes with AOI and yields an average value of zero (red circle). These phase fluctuations are corrected with relay optics, allowing a near constant visibility.}
\label{phase}
\end{figure}  

We demonstrate the robustness of the MM-TQAs against AOI variation of incoming photons by carrying out the above entanglement verification measurements for different AOIs. As shown in Fig. 4(d), for both methods, the measured visibilities are constant within experimental errors, confirming the robustness of the MM-TQAs against angular fluctuations. Due to the coupling geometry of photons into the multimode detector fiber, we vary the AOI up to 0.2$^{\circ}$. Note that this angle range is already larger than the measured error of our signal pointing system on a moving vehicle~\cite{J. -P. Bourgoin 2015}. In addition, the MM-TQA is able to recover AOI-induced phase shifts. Without correcting optics, a varying AOI also leads to phase fluctuations in the interferometer~\cite{P. B. Dixon 2009}. From our theoretical model, we anticipate a 5\,$\pi$ shift with an AOI of only $1 \times10^{-4}$ degrees [see inset of Fig. 5]. To assess the phase stability of the MM-TQA with AOI, using Method 1, we measure correlation visibilities for AOIs changing from -0.20$^{\circ}$ to +0.20$^{\circ}$ continuously over 20 sec. The measured visibilities remain almost constant within experimental errors [see Fig. 5], showing that the MM-TQA prevents AOI-caused phase fluctuations. 

\begin{figure}
\includegraphics[width=\linewidth]{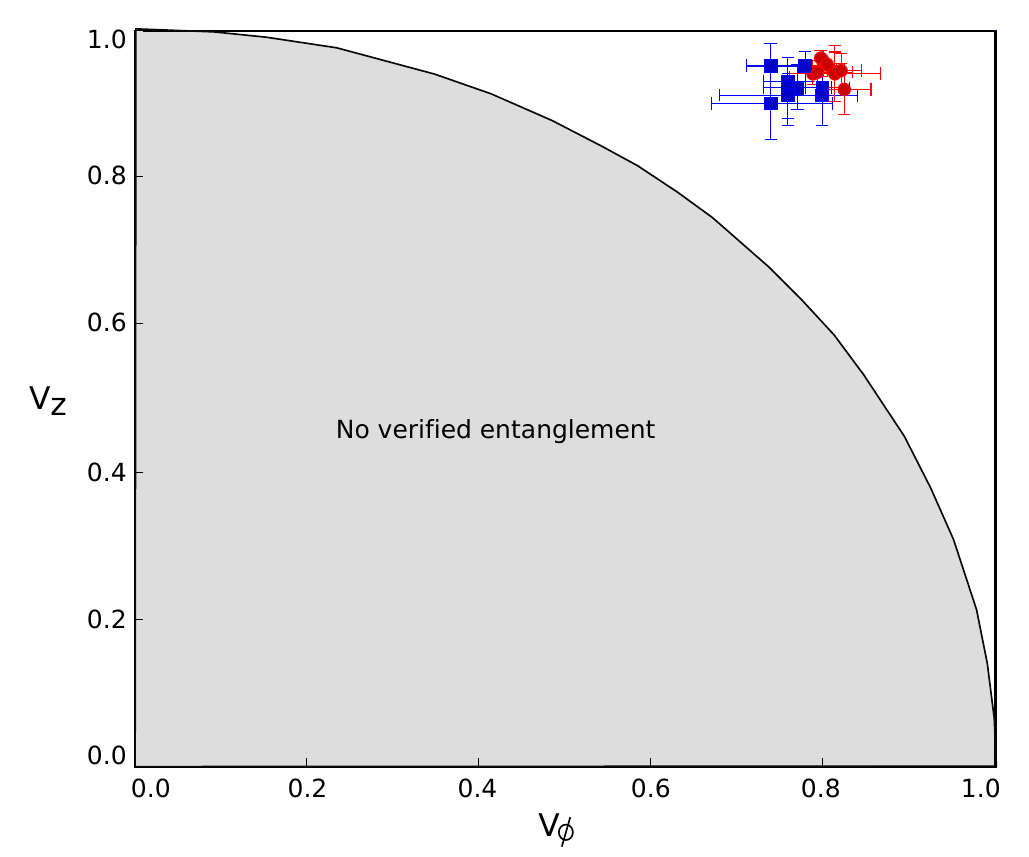}
\caption[entanglementverification]{(Color online) \textbf{Entanglement verification.} The Negative Partial Transpose (NPT) criterion~\cite{A. Peres 1996, M. Horodecki 1996} is used to obtain the required entanglement visibilities, certifying the presence of entanglement in an arbitrary 2 $\times$ 3-dimensional quantum state. Our experimental results for various angles with Method 1 (red circles) and Method 2 (blue squares) are well above the classical bound (black line).}
\label{theory}
\end{figure} 

The suitability of our MM-TQA for quantum communication is further substantiated by examining the CHSH-Bell inequality~\cite{J.F. Clauser 1969} using Method 1. We search for the maximally achievable CHSH-Bell parameter $S$ within the correlation data taken while the measurement basis of the time-bin qubits drifted slowly. The observed maximum value of $S_{\rm{exp}}$ = 2.42~$\pm$~0.05 is in good agreement with the predicted parameter $S_{\rm{theo}}$ = 2.47~$\pm$~0.02 calculated assuming the state described in Eq. (3) is transmitted through asymmetric depolarization channel [see Appendix A]. We reframe a more rigorous verification of entanglement considering the practical assumptions, including photon loss in the channel and path-dependent transmission in the MM-TQAs. We numerically find minimum visibilities required to detect the presence of entanglement for an arbitrary 2 $\times$ 3-dimensional quantum system, in which an additional dimension of the time-bin photon is taken into account due to channel transmission loss [see Appendix C]. The measured visibilities for various AOIs are found to be well above the obtained classical bound [see Fig. 6], showing clear evidence of entanglement. Note that the numerical results are valid regardless of the system efficiency and mismatched transmission between the two paths of the MM-TQAs. \\

\section{Discussion}
While theoretically a unit visibility is possible, the performance of our current MM-TQAs using imaging optics and different refractive-indexed paths show $\mathcal{V}_0$ = 0.89 $\pm$ 0.01 and 0.91 $\pm$ 0.01, respectively. These visibilities are expected to be improved considering several approaches. First, the overlap of the spatial modes in the interferometer can be improved with careful custom design and selection of optical elements and optimization of beam diameters. Second, appropriate mode-matching optics in both paths will improve the interference quality while symmetrizing dispersion at the expense of increased system complexity. Third, improved thermal stability and temperature-dependent reflective index change through dispersive optics such as a lens and a glass will minimize dispersion-induced interference degradation. Preliminary result of an updated next generation system shows a visibility of up to 0.97. 

\section{Conclusion}
We demonstrated two types of unbalanced interferometers as multimode time-bin qubit analyzers for quantum communication, which are compatible with spatially and temporally distorted photons and robust against angle of incidence fluctuations. With optical input modes emerging from a multimode optical fiber, the analyzers show an average interference visibility of up to 0.91 $\pm$ 0.01, constant with varying AOI over 0.2$^{\circ}$. The viability of the analyzers for quantum communication is substantiated by a measured entanglement visibility of up to 0.85 $\pm$ 0.01 between a polarization and a highly distorted time-bin photonic qubit. Phase stability of the analyzers is confirmed by constant correlation visibility with varying AOI. The output of the interferometers are coupled into a multimode fiber yielding a high throughput of 0.74 from input to output, mainly limited by optical surface losses. Hence, our results open the door for implementing time-bin based quantum communication experiments over multimode channels expected with moving platforms, including aircraft and satellites, or through depolarizing windows. Furthermore, recent protocols implemented in optical fiber, including coherent one-way and differential phase shift QKD protocols, could be realized over free-space channels using these analyzers. \\

\noindent Note added: After the initial release of our work on the arXiv preprint server, we became aware of related publications from T. Graham {\it et al.}~\cite{T. Graham 2015}, C. Zeitler {\it et al.}~\cite{C. Zeitler 2016}, and G. Vallone {\it et al.}~\cite{G. Vallone 2016}. The groups who authored Refs.~\cite{T. Graham 2015, C. Zeitler 2016} and \cite{G. Vallone 2016} both developed a time-bin analyzer using an unbalanced Mach-Zehnder interferometer together with imaging optics, that are conceptually similar to our Method 1. With their analyzers, the authors observed an interference visibility of 0.93 for angular variations up to 8.6 $\times10^{-3}$ degrees~\cite{C. Zeitler 2016}, and the coherent superposition of a laser pulse attenuated to the single-photon level after being reflected from a satellite~\cite{G. Vallone 2016}.

\section*{Acknowledgments}
The authors would like to thank Jacob Koenig, Rolf Horn, Evan Meyer-Scott, and Patrick Coles for useful discussions, and Martin Laforest for lending us the Hamamatsu EM-CCD camera. We gratefully acknowledge supports through the Office of Naval Research (ONR), the Canada Foundation for Innovation (CFI), the Ontario Research Fund (ORF), the Canadian Institute for Advanced Research (CIFAR), the Natural Sciences and Engineering Research Council of Canada (NSERC), and Industry Canada. 

\appendix
\section*{ Appendix A: Estimation of CHSH-Bell parameter}

The nonclassicality of time-polarization entanglement is bounded with an estimate of the Bell-CHSH inequality violation \cite{J.F. Clauser 1969}. Despite the absence of active phase control, required to set measurement bases for the time-bin qubit deterministically, we search for the maximally obtainable violation using Method 1 by varying the measurement basis. 

We first set the measurement basis for the polarization qubit to $A_1$ $\equiv$ $|\rm{Z+X}\rangle\langle \rm{Z+X}|$ using wave plates, and slowly and continuously change the path-length difference of the MM-TQA by externally heating it. This allows us to scan projection measurements for the time-bin qubit in superposition bases. Fig. 7(a) shows coincidences between a polarization qubit (two detectors, i.e., D1 and D2) and a time-bin qubit (three temporal modes). Detections in the early/late bin (middle bin) correspond to a projection of the time-bin qubit onto $B_1$ $\equiv$ $|\rm{Z}\rangle\langle \rm{Z}|$ ($B_2$ $\equiv$ $|\rm{\phi}\rangle\langle \rm{\phi}|$). Owing to the absence of the second output of the MM-TQA, we consider all possible expectation values $E(A_{\rm{i}}, B_{\rm{j}})$ between any two points in time, i.e., $t_1$ and $t_2$ [see Fig. 7(b)], which is defined as

\begin{eqnarray}
E(A_{\rm{i}}, B_{\rm{j}})&=&\frac{{N}_{\rm{ij}}^{++}+{N}_{\rm{ij}}^{--}-{N}_{\rm{ij}}^{+-}-{N}_{\rm{ij}}^{-+}}{{N}_{\rm{ij}}^{++}+{N}_{\rm{ij}}^{--}+{N}_{\rm{ij}}^{+-}+{N}_{\rm{ij}}^{-+}}. \label{eq:expectation} 
\end{eqnarray}

\noindent Here, $\rm{N}_{\rm{ij}}$ are the coincidence counts for the projections $A_{\rm{i}} \otimes B_{\rm{j}}$, where i, j $\in$\{1,2\} and superscript($+,-$) denotes two outcomes of the projection measurement. Among all the computed expectation values, we find the absolute maximum expectation value. We then change the measurement basis for the polarization qubit to $A_2$ $\equiv$ $|\rm{Z-X}\rangle\langle \rm{Z-X}|$ and repeat the procedure. Finally, we compute the CHSH-Bell inequality parameter

\begin{equation}
S=|E(A_1, B_1)-E(A_1, B_2)+E(A_2, B_1)+E(A_2, B_2)|
\label{eq:CHSH-Bellparameter}
\end{equation}

\noindent and find the value of $S_{\rm{exp}}$\,=\,2.42 $\pm$ 0.05, which is clearly above classical bound $S$\,=\,2. To see whether this value agrees with the measured visibilities, we model the two-qubit state with noise, described by an asymmetric depolarization channel, on a time-bin qubit. The output state is described by 

\begin{equation}
\rho_{\rm{out}} = (1- \sum_{\rm{j=X,Y,Z}}p_{\rm{j}}) |\psi\rangle\langle \psi| +  \sum_{\rm{j=X,Y,Z}} p_{\rm{j}} (\mathbb{I} \otimes \rm{j}) |\psi\rangle\langle \psi| (\mathbb{I} \otimes \rm{j}), 
\end{equation}

\noindent where $p_{\rm{j}} \,(\rm{j=X,Y,Z})$ and $\rm{j}$ denote the depolarization probability and single-qubit Pauli operator, respectively. Here, $|\psi\rangle$ is the input state, described in Eq. (3). Assuming unbiased depolarizations in superposition bases, i.e., $(p_{\rm{X}} = p_{\rm{Y}}) \equiv p_{\rm{\phi}}$, we calculate expectation values $E(A_{\rm{i}}, B_{\rm{j}})$ = Tr $(\rho_{\rm{out}} \enskip A_{\rm{i}} \otimes B_{\rm{j}}) $ for given measurement bases. Using the definition of visibility, we further represent the CHSH-Bell parameter $S_{\rm{theo}}$ = $\sqrt{2} (\mathcal{V}_{\rm{Z}} + \mathcal{V}_{\rm{\phi}})$ as a function of entanglement visibilities. We find $S_{\rm{theo}}$ = 2.47 $\pm$ 0.02, in accordance with our measured value $S_{\rm{exp}} = 2.42 \pm 0.05$. 

\begin{figure*}

\includegraphics[width=\textwidth]{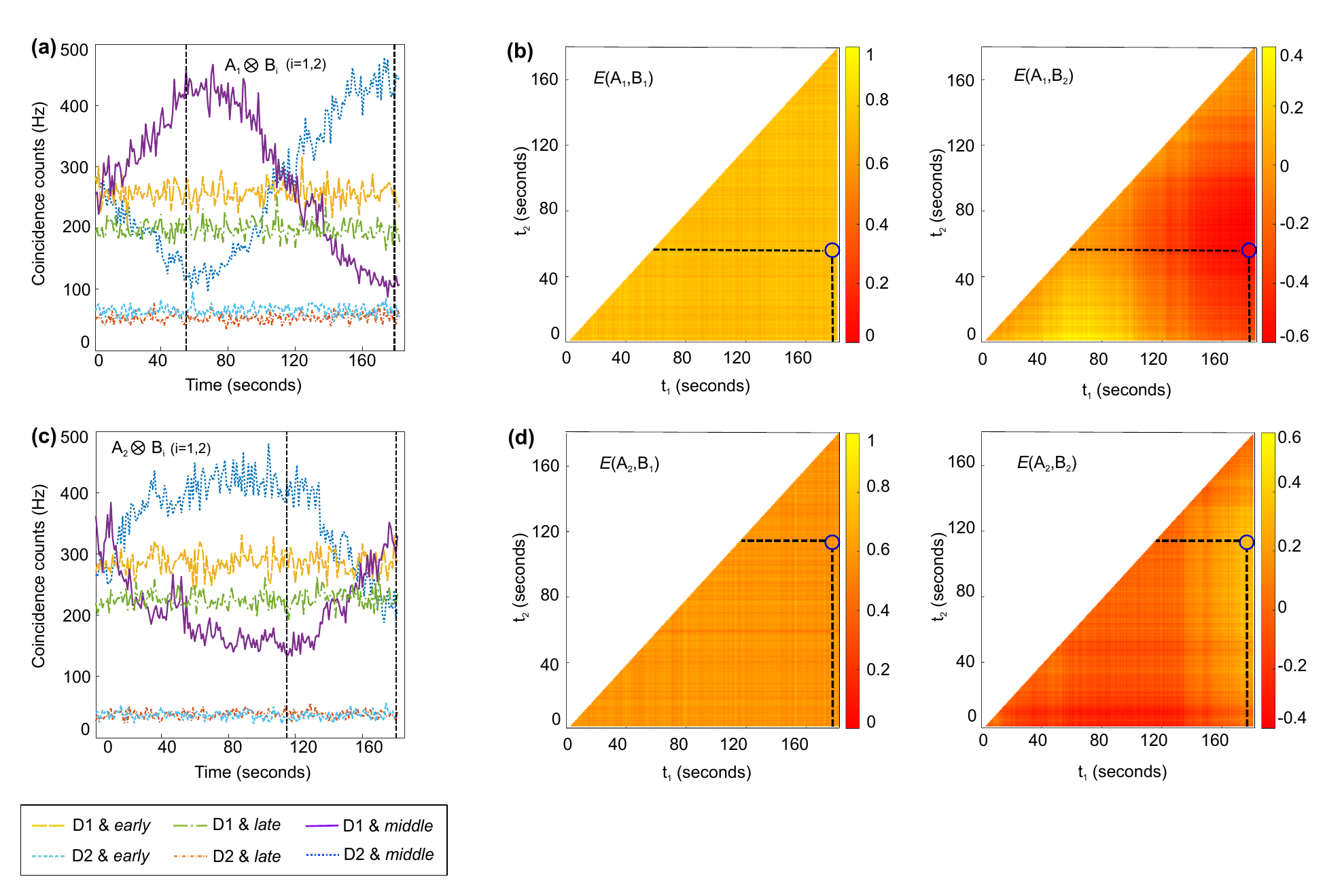}
\caption[CHSH-Bellinequality]{(Color online) \textbf{Estimation of the CHSH-Bell parameter.} \textbf{(a)} Long-dashed yellow and short-dashed light blue lines are coincidences from joint projections $\rm{(Z+X)}\otimes\rm{+Z}$ (early temporal bin), long dash-dotted green and short dash-dotted orange lines from $\rm{(Z+X)}\otimes\rm{-Z}$ (late temporal bin), and solid purple and dotted blue lines from $\rm{(Z+X)}\otimes\rm{\phi}$ (middle temporal bin). Dashed black lines are times at which maximal expectation values are extracted. \textbf{(b)} Surface plot of calculated expectation values for the projections in (a) between any two points in time. \textbf{(c)} Long-dashed yellow and short-dashed light blue lines are coincidences from joint projections $\rm{(Z-X)}\otimes\rm{+Z}$, long dash-dotted green and short dash-dotted orange lines from $\rm{(Z-X)}\otimes\rm{-Z}$, and solid purple and dotted blue lines from $\rm{(Z-X)}\otimes \rm{\phi}$. Dashed black lines are times at which maximal expectation values are extracted. \textbf{(d)} Surface plot of calculated expectation values for the projection measurements in (c). The Bell-CHSH parameter is calculated using the maximum expectation values. The measurement duration is chosen arbitrarily and yields a violation of the inequality of 2.42 $\pm$ 0.05.}
\label{nonlocality}
\end{figure*}

\section*{Appendix B: Long-term phase stability of our MM-TQA}

Our MM-TQAs are passively stabilized by enclosing them with black cardboard. In order to assess the phase stability of the MM-TQA, using Method 1, we perform joint-projection measurements onto superposition bases more than a half hour. The time-bin qubit is projected onto $|\rm{\phi}\rangle\langle \rm{\phi}|$ and the polarization qubit alternatively between $|\rm{\phi^{\prime}}\rangle\langle\rm{\phi^{\prime}}|$ or $|\rm{\phi^{\prime}+\pi/2}\rangle\langle\rm{\phi^{\prime}+\pi/2}|$. As shown in Fig.~8, the correlation visibility  remains always higher than 0.65, which is well above the required value for verifying entanglement, given entanglement visibility $\mathcal{V}_{\rm{Z}}$~=~0.95 $\pm$ 0.01 [see Fig. 6].

\section*{Appendix C: Entanglement verification}

\noindent The ability to verify effective entanglement is a necessary condition for secure QKD~\cite{M. Curty 2004}. This is especially important in the absence of a complete security analysis of a QKD implementation, and applies to prepare-and-measure QKD as well as entanglement-based schemes. We assume that the 
spontaneous parametric down-conversion process generates a pair of 
photons with negligible multiple-photon-pair events.  
Each photon is a polarization qubit and the pair of photons is 
potentially entangled. By detecting a photon A in the pair, we can 
herald the other photon B.  After the conversion 
from polarization qubit to time-bin qubit, the photon B
is transmitted to the MM-TQA. Suppose that 
Alice holds the polarization qubit while Bob holds the time-bin qubit. 
To include conversion and transmission losses of the time-bin qubit,
we enlarge the dimension of Bob's system from $2$ to $3$ by
adding a dimension corresponding to no photon arriving at Bob. Hence,
the final state $\rho$ shared by Alice and Bob is a $2 \times 3$-dimensional 
state. We need to verify whether or not the state $\rho$ is entangled using 
only the measurement results in $\mathcal{V}_{\rm{Z}}=0.95 \pm 0.01\,(0.92 \pm 0.01)$ and $\mathcal{V}_{\rm{\phi}}=0.80 \pm 0.01\,(0.77 \pm 0.01)$ with Method 1(2) without further assumptions on the state. Since the measurements of Alice and Bob are block-diagonal with respect to the subspaces of total photon number, as we will
show below in Eq.~(7) and Eq.~(8), we can also assume 
without loss of generality that the state $\rho$ shows the same structure. This 
follows from the fact that the measurement structure allows us to assume that a 
quantum nondemolition measurement of the total photon number is executed before 
the actual measurement itself.

\begin{figure}
\includegraphics[width=\linewidth]{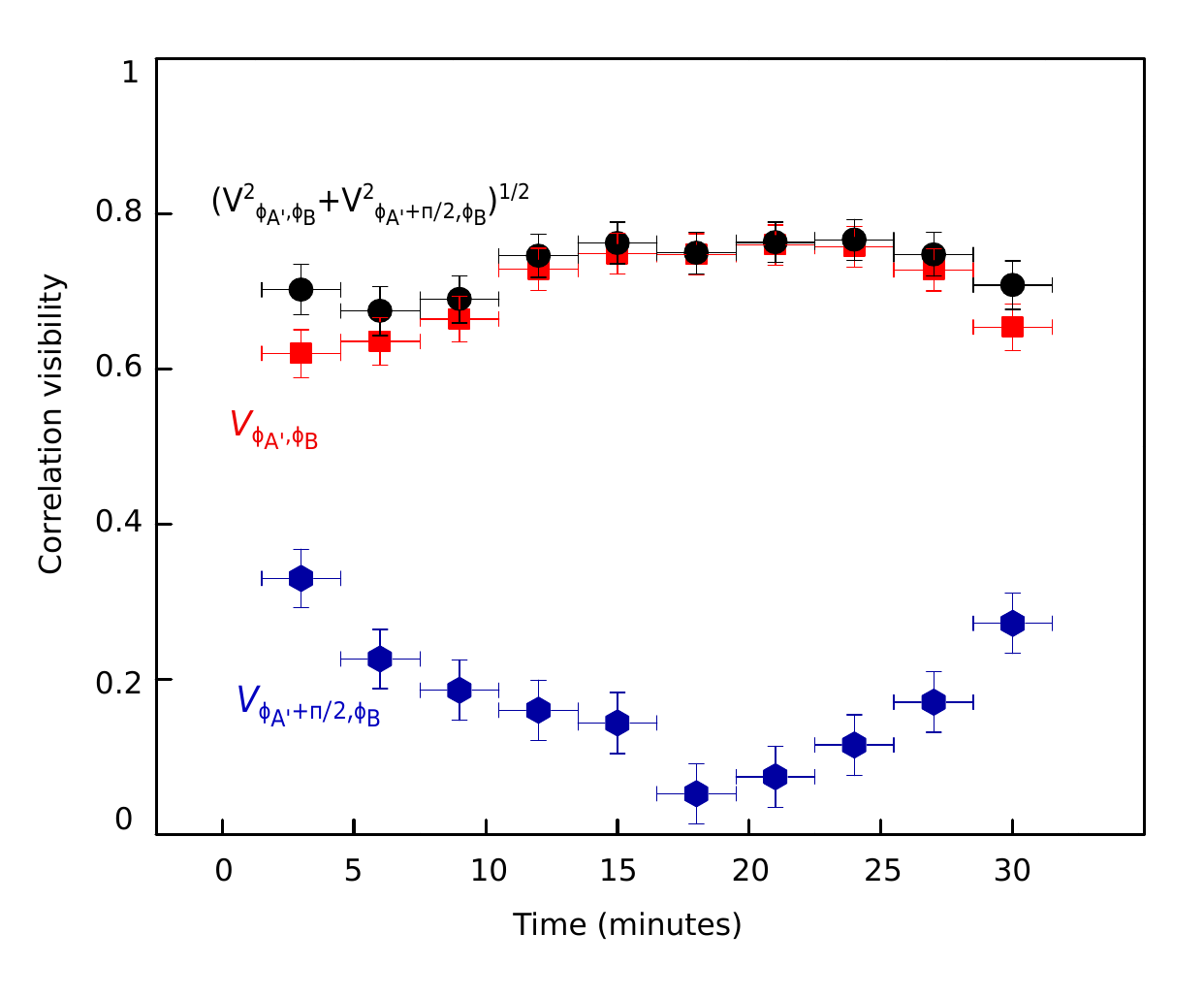}
\caption[expectationtime]{(Color online) \textbf{Long-term phase stability of our MM-TQA.} Red squares are expectation values for projection measurements $\rm{\phi_{A^{\prime}}} \otimes \rm{\phi_B}$ ($V_{\phi_{A^{\prime}}, \phi_B}$) and blue hexagons for $\rm{\phi_{A^{\prime}}+\pi/2} \otimes \rm{\phi_B}$ ($V_{\phi_{A^{\prime}}+\pi/2, \phi_B}$). Each measurement is averaged over 3 min. Black circles are average correlation visibility values $\sqrt{V_{\phi_{A^{\prime}}, \phi_B}^2+V_{\phi_{A^{\prime}}+\pi/2, \phi_B}^2}$. The phase drifts slowly, on the order of $\pi$/2 over an half an hour, showing the stability of our MM-TQA. The average expectation value is always higher than the required expectation value for entanglement verification.}
\label{stability}
\end{figure}

In order to verify entanglement, we need to know how to accurately describe 
the measurements on the polarization and the time-bin qubit. 
For the polarization qubit, we measure it in the horizontal/vertical or 
diagonal/anti-diagonal basis, i.e. along the Z- or X-axis in the 
Bloch sphere. These measurements are represented as  

\begin{multline}
M_{\rm{H}}= \Bigg[
\begin{array}{cc}
1\,0 \\
0\,0 
\end{array}
\Bigg]\,\,\,
M_{\rm{V}}= \Bigg[
\begin{array}{cc}
0\,0 \\
0\,1 
\end{array}
\Bigg]\,\,\,\\
M_{\rm{D}}= \frac{1}{2} \Bigg[
\begin{array}{cc}
1\,1 \\
1\,1 
\end{array}
\Bigg] \,\,\,
M_{\rm{A}}= \frac{1}{2} \Bigg[
\begin{array}{cc}
1\,-1 \\
-1\,\,\,\,1 
\end{array}
\Bigg],
\end{multline}

\noindent where the subscript indicates measurement outcome, and H, V, D, or A 
denotes the horizontal, vertical, diagonal, or anti-diagonal polarization. 
On the other side, for the time-bin qubit, the photon loss in the 
long path or the short path of the MM-TQA could be different from each other. 
Hence, the operators corresponding to measurement of the time-bin qubit in the 
early/late basis or in the superposition bases could deviate from the ideal case. Without loss of generality, we can choose the relative phase between the early- and late-basis states 
in the superposition basis to be zero. Therefore, in the basis in which
the basis states are no photon, one photon in the early bin and 
one photon in the late bin, these measurements can be written as 

\begin{eqnarray}
&M_{\rm{E}}= \frac{1}{4}\left[
\begin{array}{ccc}
0&0&0 \\
0&\eta_{\rm{S}}&0 \\
0&0&0
\end{array}
\right] \quad
M_{\rm{L}}= \frac{1}{4} \left[
\begin{array}{ccc}
0 & 0 & 0 \\
0 & 0 & 0 \\
0 & 0 & \eta_{\rm{L}}
\end{array}
\right]\\
&M_{\rm{X}}= \frac{1}{4} \left[
\begin{array}{ccc}
0 & 0 & 0 \\
0 & \eta_{\rm{L}} & \sqrt{\eta_{\rm{S}} \eta_{\rm{L}}} \\
0 & \sqrt{\eta_{\rm{S}} \eta_{\rm{L}}} & \eta_{\rm{S}}
\end{array}
\right]\quad 
M_{\emptyset}= I-M_{\rm{E}}-M_{\rm{L}}-M_{\rm{X}},\nonumber
\end{eqnarray}

\noindent where the subscript E, L, X, or $\emptyset$ means that the measurement outcome
is early time, late time, the superposition of the early and late time, or no detection, respectively. No-detection events are due to detection inefficiency 
and the absence of the second output in the MM-TQA. In Eq.~(8), 
$\eta_S$ or $\eta_L$ is the respective transmission efficiency in the short path
or the long path of the MM-TQA. Note that in our experiment $\eta_S$ and 
$\eta_L$ are very close to each other. 

After knowing the description of Alice's and Bob's joint state $\rho$ and that of 
their measurements, we can verify entanglement by the negative partial-transpose (NPT) 
criterion~\cite{A. Peres 1996}. The NPT criterion is used because this criterion is satisfied if and only if a state is entangled, given the 
state is $2\times 2$- or $2 \times 3$-dimensional~\cite{M. Horodecki 1996}. The NPT 
criterion has been applied to verify entanglement in QKD systems, such as  
in Ref.~\cite{M. Curty 2007} .  Explicitly, we verify entanglement by solving the following 
semidefinite program (SDP): finding $\rho$ subject to $\rho \geq 0, \, Tr(\rho)=1,\,  \rho^\Gamma \geq 0$ that satisfies 

\begin{equation}
\begin{array}{cc}
Tr[\rho(M_{\rm{H}}\otimes M_{\rm{E}}-M_{\rm{V}}\otimes M_{\rm{E}})]\\
=\mathcal{V}_{\rm{+Z, -Z}}Tr[\rho(M_{\rm{H}}\otimes M_{\rm{E}}+M_{\rm{V}}\otimes M_{\rm{E}})] \\
\\
Tr[\rho(M_{\rm{V}}\otimes M_{\rm{L}}-M_{\rm{H}}\otimes M_{\rm{L}})]\\
=\mathcal{V}_{\rm{+Z,-Z}} Tr[\rho(M_{\rm{V}}\otimes M_{\rm{L}}+M_{\rm{H}}\otimes M_{\rm{L}})] \\
\\
Tr [\rho(M_{\rm{D}}\otimes M_{\rm{X}}-M_{\rm{A}}\otimes M_{\rm{X}})]\\
=\mathcal{V}_{\rm{\phi}}\ Tr [\rho(M_{\rm{D}}\otimes M_{\rm{X}}+M_A\otimes M_{\rm{X}})],
\end{array}
\end{equation}

\noindent where $\Gamma$ is the partial-transpose operation on a subsystem, such as on the 
polarization-qubit subsystem, and $\otimes$ denotes the tensor product.  Note that,
we formulate the last three constraints according to the measured visibilities.
The first two are based on entanglement visibilities $\mathcal{V}_{\rm{+Z, +Z}}$ 
and $\mathcal{V}_{\rm{-Z, -Z}}$ conditioned on measurement outcomes of the time-bin qubit 
being early time and late time, respectively. The last constraint 
is based on entanglement visibility $\mathcal{V}_{\rm{\phi}}$, where the 
time-bin qubit comes out in the middle bin. Since the MM-TQA has only one output, we cannot 
differentiate the case when the photon comes out from the second output if this 
output exists from the case when the photon is lost over the transmission. Hence, 
we cannot formulate two constraints based on $\mathcal{V}_{\phi}$.

In our experiment, we verified that within experimental errors 
the visibilities $\mathcal{V}_{\rm{+Z, +Z}}=\mathcal{V}_{\rm{-Z,-Z}}$. So, for solving the SDP 
in Eq.~(9) we set $\mathcal{V}_{\rm{+Z,+Z}}=\mathcal{V}_{\rm{-Z,-Z}}=\mathcal{V}_{\rm{Z}}$. 
Using the measured results of Method 1(2) $\mathcal{V}_{\rm{Z}}=0.95 \pm 0.01\,(0.92 \pm 0.01)$ and $\mathcal{V}_{\rm{\phi}}=0.80 \pm 0.01\,(0.77 \pm 0.01)$, the SDP 
in Eq.~(9) is not feasible, signifying that the state $\rho$ 
must be entangled. Furthermore, by numerically checking over which values of 
$\mathcal{V}_{\rm{Z}}$ and $\mathcal{V}_{\rm{\phi}}$ the SDP 
in Eq.~(9) is not possible,
we are able to upper bound the required visibilities $\mathcal{V}_{\rm{Z}}$ and $\mathcal{V}_{\rm{\phi}}$ 
that certify the presence of entanglement in the system. The numerical 
results are shown in Fig. 6. From this figure, one 
can see that our visibility result at any observed incident angle witnesses entanglement 
with high confidence. Finally, we would like to note two points.  First, 
the constraints considered in Eq.~(9) are independent of the 
transmission or conversion loss of the photon arriving at the MM-TQA, and even 
independent of the common photon loss in the two different paths of the 
MM-TQA. Therefore, the upper bounds on the visibilities $\mathcal{V}_{\rm{Z}}$ and $\mathcal{V}_{\rm{\phi}}$ obtained for verifying entanglement are independent of all of these different losses.
Second, our obtained classical boundary [see Fig.6] is even independent of the relative loss between the 
two paths of the MM-TQA. 
 
\bibliographystyle{apsrev4-1}

\end{document}